# Calculations of Real-System Nanoparticles Using Universal Neural Network Potential PFP

Gerardo Valadez Huerta[a,*], Yusuke Nanba[a], Iori Kurata[c,*], Kosuke Nakago[c], So Takamoto[c], Chikashi Shinagawa[c], Michihisa Koyama[a,b,*]

[a] Research Initiative for Supra Materials, Shinshu University

[b] Open Innovation Institute, Kyoto University

[c] Preferred Networks, Inc.

*valadez@shinshu-u.ac.jp;  kurata@preferred.jp;  koyama_michihisa@shinshu-u.ac.jp, koyama.michihisa.5k@kyoto-u.ac.jp

It is essential to explore the stability and activity of real-system nanoparticles theoretically. While applications of theoretical methods for this purpose can be found in literature, the expensive computational costs of conventional theoretical methods hinder their massive applications to practical materials design. With the recent development of neural network algorithms along with the advancement of computer systems, neural network potentials have emerged as a promising candidate for the description of a wide range of materials, including metals and molecules, with a reasonable computational time. In this study, we successfully validate a universal neural network potential, PFP, for the description of monometallic Ru nanoparticles, PdRuCu ternary alloy nanoparticles, and the NO adsorption on Rh nanoparticles against first-principles calculations. We further conduct molecular dynamics simulations on the NO-Rh system and challenge the PFP to describe a large, supported Pt nanoparticle system.

1. **Introduction**

The computational design of catalysts is a promising field in current development [1–4]. Highly challenging catalysts are based on monometallic or multinary alloy nanoparticles [5–10]. Two imperative issues are posed to the computational design of multinary alloy nanoparticles [11]. One is the stability, and the other is the heterogeneities of the real-system catalysts, both of which should be considered to realize a practical design of catalysts [11]. Parallel-computing first-principles calculations are adopted to incorporate heterogeneous structures of nanoparticles and discuss their stabilities [12]. In general, nanoparticle catalysts are generally synthesized on support materials [13]. Despite the existing DFT studies of supported nanoparticles [14–16], the expensive computational cost of the first-principles calculations makes it challenging to tackle real-system catalysts. Moreover, the reaction dynamics under a finite operation temperature are often different from the reaction pathway analyzed by static transition-state calculations. Therefore, it is expected to develop a method for elucidating the stability and activity of the real-system catalyst with a reasonable computational cost and accuracy.

Classical force fields, such as OPLS-AA [17] or COMPASS [18], provide a possibility to fill this gap. While classical force fields can be applied to the discussion of the stability of the catalysts, they require the first-principles calculations to determine the force field parameters; thus, they lack the predictability toward new material systems. Further, they cannot describe the chemical reaction by its nature. Reactive force fields have been developed, such as ReaxFF by Van Duin *et al.* [19], to deal with chemical reactions. Its effectiveness for investigating the reaction dynamics considering the real-system structure is reported [20]. However, improving the accuracy of ReaxFF is a big hurdle.

Recent advancement of deep learning and computational hardware has led to the emergence of neural network potentials (NNP) [21–23], which can simultaneously support high accuracy, reasonable computational costs and chemical reaction dynamics. Its application to practical materials design is expected in many fields. However, an existing NNP's limitation is the elements and material types supported by the software. For example, the latest version of the ANI NNP, the ANI-2x [24], only supports seven elements (H, C, N, O, F, Cl, S), which undoubtedly allows it to be used for the description of a wide range of organic molecules, but not of metallic systems. In addition, the Open Catalyst Project [4], which targets molecular adsorption on slab models, does not apply to nanoparticle systems.

In the previous study [25], a universal neural network potential, PFP, was developed, supporting a wide range of elements and various material types with the accuracy comparable to first-principles calculations. This work uses the PFP version 0, which does not explicitly contain nanoparticle systems in training datasets. We show the application of the PFP to various real-system nanoparticles. Further, the PFP is applied to the dynamics of an adsorbate on a nanoparticle and a supported nanoparticle system.

## 2. Computational Details

### 2.1 Neural Network Potential

The PFP is an already trained model and the details can be found in the preceding paper [25]. Here, we briefly introduce the characteristics of the PFP.

Model Description

The architecture of PFP is based on the Graph Neural Network, whereas E(3) invariance is implemented in the architecture, which prevents the introduction of artificial errors that break the physical symmetry with respect to the atomic positions. The system's total energy is defined as the sum of the energies of each atom subjected to local interactions resulting in an extensive property. The local interactions are guaranteed to disappear when the distance between the nearest atoms exceeds a cutoff radius of 6 Å. For example, in the computational model of nanoparticles under periodic boundary conditions, the vacuum layer will not self-interact beyond the unit cell if it is larger than this distance. Moreover, the energy surface is designed to be differentiable to higher orders depending on the atoms' position. Therefore, there is no discontinuity in the energy or force even before or after the cutoff distance, and it can be used together with various finite difference methods.

In addition to energy and force, PFP predicts the Bader charge as a physical quantity per atom [26]. The value of charge changes depending on the arrangement of atoms. It can be said to be a polarizable force field as defined by Baker [27]. Note that the total charge of the entire system is constrained to be neutral. Also, the response to external fields, such as electric fields, cannot be calculated in the present version.

Dataset Description

The dataset for training the PFP has been constructed by density functional theory (DFT) calculations. The data collection has been designed to deal with various materials [25]. The PFP crystal dataset covers various systems, such as bulk, slab, cluster, and molecule. The number of atoms in clusters in the dataset is smaller than the ones in nanoparticles represented in this study. The dataset was computed using the Vienna *ab initio* simulation package (VASP) [28–31] with the projector-augmented wave [32, 33] method. The Perdew-Burke-Erzernhof functional with the generalized gradient approcoximation [34] was used to describe the exchange and correlation interactions. The cutoff energy was set to be 520 eV, and Gaussian smearing was adopted with a smearing width of 0.05 eV. The *k*-point sampling is a Monkhorst-Pack grid [35] of 1000 *k*-points / reciprocal atom proportional to the reciprocal length of the *a*, *b*, and *c* axes for bulk and slab systems. Clusters and molecules were calculated with vacuum regions, and only Γ points were sampled for them.

To investigate the applicability of PFP to the real-system nanoparticles, we referred to the DFT calculations published in the literature [36–38]. These conditions are not the same as those of the training sets, but are not significantly different, thus being reliable.

**2.2 Theoretical Methods and Calculation Conditions**

The definitions for structural and thermodynamic properties used in the next section should be summarized at first. For monometallic nanoparticles, the cohesive energy $\varepsilon_{\text{coh}}$ given by

$$\varepsilon_{\text{coh}} = \frac{E_{\text{tot}} - N \cdot \varepsilon_0}{N} \tag{1}$$

is used as a measure for their stability, as in the work by Nanba *et al.* [37]. Here, $E_{\text{tot}}$ is the total energy of the nanoparticle, $N$ the number of atoms in the nanoparticle and $\varepsilon_0$ is the ground state energy of a single atom.

Pairwise multicomponent short-range order (SRO) [39] was used to evaluate the homogeneity of the solid-solution state for ternary alloy nanoparticle systems. The definition of the SRO is the same as that of the Warren-Cowley parameter:

$$\alpha_{\text{AB}} = 1 - \frac{p_{\text{AB}}}{x_{\text{B}}}, \tag{2}$$

whereas, $p_{\text{AB}}$ is the probability of finding an atom of type B as the nearest neighbor of an atom of type A and $x_{\text{B}}$ is the mole fraction of the B atoms in the system. The binary WC parameter becomes zero when the probability $p_{\text{AB}}$ is equal to the mole fraction $x_{\text{B}}$. When a local order of neighbor atoms B is built, the probability $p_{\text{AB}}$ becomes lower or higher than $x_{\text{B}}$ and the WC parameter becomes lower or higher than zero.

To analyze the stability of alloy nanoparticles, the excess energy was calculated as follows:

$$\varepsilon_{\text{excess}} = \frac{1}{N}\left(E_{\text{tot}} - \sum_{k=1}^{K} x_{k} \cdot E_{\text{tot},k}\right), \tag{3}$$

whereas $E_{\text{tot}}$ is the total energy of the alloy nanoparticle, $x_k$ is the mole fraction of the k alloy's component, $E_{\text{tot},k}$ is the total energy of each k monometallic nanoparticle, and $N$ is the total number of atoms.

Finally, the adsorption energy $\varepsilon_{\text{ads}}$ for a molecule adsorbed to a nanoparticle is calculated as

$$\varepsilon_{\text{ads}} = E_{\text{tot,NP-M}} - E_{\text{tot,NP}} - E_{\text{tot,M}}, \tag{4}$$

whereas $E_{\text{tot,NP-M}}$ is the total energy of the whole optimized system, including the nanoparticle and the adsorbed molecule, $E_{\text{tot,NP}}$ is the total energy of the optimized nanoparticle and $E_{\text{tot,M}}$ is the total energy of the optimized single molecule.

In some of the studied cases for validation, no DFT data is available in the literature. Therefore, we use mainly the data of the research group of M. Koyama [36, 37]; we applied the same methods reported in those studies to produce our DFT data for the cases that were not available in the literature. The methods to produce this extra data differentiate from the methods used to generate the dataset by a cutoff energy of 400 eV. Furthermore, all calculations were independently performed within the research group of M. Koyama.

Throughout this study, we performed optimization procedures using the PFP. The standard Broyden-Fletcher-Goldfarb-Shanno (BFGS) method [40, 41] is used for all optimization procedures as it is implemented in ASE [42] with a force tolerance of 0.001 eV/Å. While the lattice size and the position of the atoms are optimized for the simple lattices, only the position of the atoms is optimized for the nanoparticle systems. Computational conditions that differ from those described here, such as using the FIRE algorithm [43] for optimization or other force tolerance, are indicated in the corresponding section.

All MD simulations are performed in this study using ASE [42]. We use the optimized structures as start configurations and set the initial velocities using a Maxwell Boltzmann Distribution at the specified temperature. We use the Velocity-Verlet algorithm for time integration with a time step of 0.5 fs. The simulations are performed with periodic boundary conditions in the canonical ensemble. To realize this, we use the NVT Nosé-Hoover thermostat implemented in ASE with a damping factor for the temperature a hundred times larger than the time step. The simulation time is 20000 steps, whereas 10000 steps were enough to achieve equilibration for all simulations performed here. The error of all ensemble averages is calculated with Friedberg and Cameron's [44] standard block average method.

## 3. Results and Discussion

### 3.1 Validation of the Neural Network Potential

Four different systems are calculated to validate the PFP and show its wide range of applications. The results are compared with our DFT results or DFT results from rigorous studies from the literature. First, we discuss the cohesive energy of monometallic Ru nanoparticles, comparing the results with those of Nanba *et al.* [37]. Then the excess energy of ternary PdRuCu alloy by PFP is compared with that of DFT. The interaction of NO molecule on $Rh_{201}$ is next investigated, referring to the results of Nanba and Koyama [36]. Furthermore, we analyze the structure and phase stability of metallic nanoparticles similar to the work by Rivera Rocabado *et al.* [38].

Ru Nanoparticles

To show the capability of the PFP for the description of monometallic real-system nanoparticles, we applied PFP to the results of Nanba *et al.* [37] for the cohesive energy for Ru nanoparticles with different structures (5-fold twinned decahedral fcc, icosahedral fcc, truncated octahedral fcc and hcp) for different nanoparticle sizes. For this purpose, we first optimized the unit Ru fcc and hcp lattice using the PFP and then optimized the different nanoparticle structures with different sizes up to 1000 atoms. The FIRE algorithm [43] was used for optimization. The cohesive energy for the nanoparticles was calculated using Eq. (1) with $\varepsilon_0 = 0$, because the ground state energy of a single Ru atom in classical MD is zero. We repeated the simulation with the EAM potential of Fortini *et al.* [45]. Unfortunately, it was only possible to optimize the atoms' positions in the simple lattice using the EAM potential this time because ASE lacks the calculation of the stress-tensor using this potential.

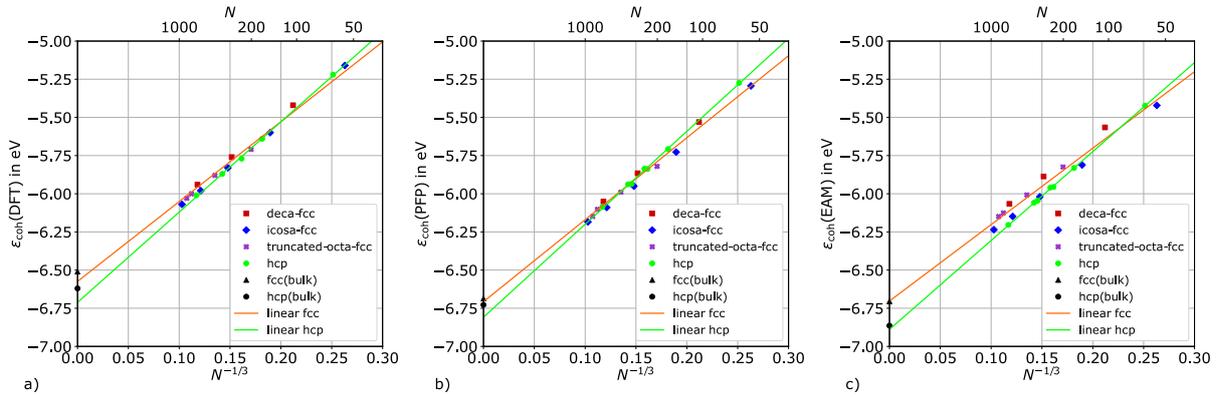

**Figure 1** Calculated cohesive energies $\varepsilon_{coh}$ for all Ru nanoparticle structures investigated here: a) 5-fold twinned decahedral fcc (in red), b) icosahedral fcc (in blue), c) truncated-octahedral fcc (in purple), and the hcp (in green) Ru structures. The values are plotted in dependency on the nanoparticle size $N^{-1/3}$. a) Values from the DFT study of Nanba *et al.* [37]. b) Values calculated using the PFP and c) Values calculated using the EAM potential of Fortini *et al.* [45].

In Figure 1, the values for $\varepsilon_{coh}$ are given in dependency on the particle size $N^{-1/3}$. We plotted the values from the DFT study of Nanba *et al.* [37] in diagram a), the calculated values using the PFP in diagram b) and the calculated values using the EAM potential in diagram c). All values show a linear dependency to $N^{-1/3}$. To further analyze the accuracy of the progression data depending on the particle size, we conducted two linear regressions with the values for all fcc structures and the values for the hcp structure. The resulting linear regressions are given along with the corresponding values in Figure 1 for the calculation using the PFP and EAM potential. The linear regression resulting from the values for the fcc structures led to a slope of 5.37(13) using the PFP and to a slope of 5.00(22) using the EAM potential. The slope calculated from the DFT values was of 5.22(16). Despite that all values were equivalent within the statistical uncertainty, the progression resulting from the calculations with the PFP reproduced the DFT progression more effectively with $R^2$ value of 0.994 similar to the DFT value of 0.990 in comparison to the $R^2$ value of 0.975 resulting from the EAM calculation. For the hcp structure, the resulting slope values were 6.08(10), 5.826(52), and 5.913(97) for calculations using the PFP and the EAM potential as well as the DFT values from Nanba *et al.* [37], respectively. Only the slopes between the PFP and the DFT values are statistical equivalent, whereas all linear regressions result in $R^2$ values of 0.999. From this discussion, we concluded that the values calculated using the PFP showed higher accuracy than those by the EAM. Further discussion can be found in the supporting information.

Stability of (PdRuCu)$_{201}$ ternary alloy nanoparticles

The PFP was tested further to describe the stability of (PdRuCu)$_{201}$ ternary alloy nanoparticles. This system showed the best performance for the PFP to predict the stability of alloy nanoparticles among the various other binary or ternary alloy nanoparticle systems investigated. We decided, thus, to summarize exemplarily the results for this ternary system here. Truncated octahedral (PdRuCu)$_{201}$ nanoparticles with different compositions Pd$_z$Ru$_z$Cu$_{(201-2z)}$ ($z = 67, 80, 90$) and configurations (solid solution (SS), segregation (SG), Cu-vertex (VT)) were investigated (see Figure S2, supporting information for details). We analyzed 26 SS (nine for $z = 67$, eleven for $z = 80$ and six for $z = 90$), twelve VT (all for $z = 90$), and two segregation (for $z = 67$ and $z = 80$) configurations. DFT calculations were also conducted for each configuration using the methods described in section 2 to check the accuracy of the PFP describing these systems.

The resulting mean values of the SRO $\alpha_{PdRu}$, $\alpha_{PdCu}$ and $\alpha_{RuCu}$ for each configuration of the (PdRuCu)$_{201}$ systems can be found in

Table **S1** of the supporting information. All SRO's of SG configurations showed values much higher or lower than zero, and those of SS were confirmed to be close to zero, i.e., randomized. For the Cu-VT structures, the SRO's were also random for the Pd and the Ru. However, $\alpha_{PdCu}$ and $\alpha_{PdRu}$ were higher than zero because of the specified configurations of Cu occupying vertex sites, where the coordination number is the smallest, leading to a lower probability to find Cu atoms as nearest neighbor.

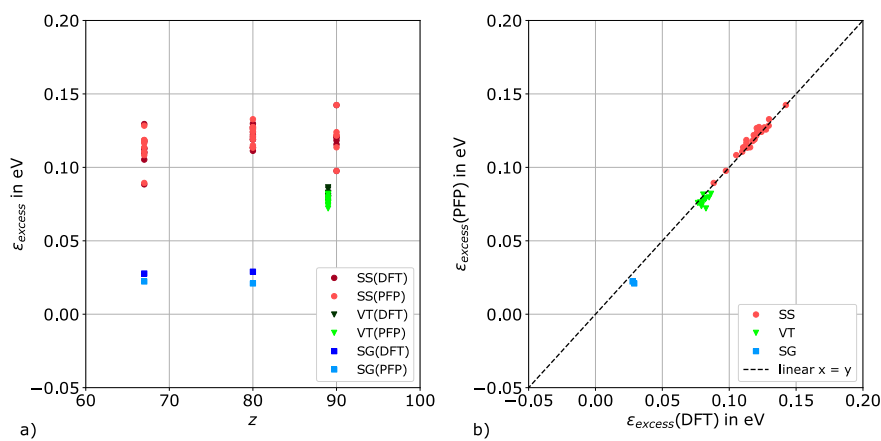

**Figure 2** a) Calculated excess energy $\varepsilon_{excess}$ using PFP (light colors) and DFT (dark colors) and b) comparison between both approaches for different Pd$_z$Ru$_z$Cu$_{201-2z}$ nanoparticles in dependency on the variable $z$. Red circles: solid solution (SS). Green triangles: Cu-vertex (VT). Blue squares: segregation (SG).

In Figure 2a, the excess energy is given for all Pd$_z$Ru$_z$Cu$_{201-2z}$ in dependency on the variable z, which is sufficient to describe the structures provided here. We could not observe any substantial dependency of the nanoparticle stability from the DFT values on the alloy composition but on the structures. The most unstable structure was the SS, followed by the VT and the SG. Not only was this behavior predicted effectively by the PFP, but also each DFT value was well-reproduced, as it can be seen from Figure 2b. Moreover, the calculated values are distributed close to the identity line for all nanoparticle configurations with an $R^2$ value of 0.990. It should be noted, that the PFP underestimated the values for the SG structures. However, the absolute values for the absolute deviations ranged for all structures between $7.5 \cdot 10^{-5}$ eV and $0.011$ eV (see supporting information, Figure S3). This is a high level of accuracy accounting for the small order of magnitude of the excess energies, the values of which laid between $0.0210$ eV and $0.1424$ eV.

Adsorption of NO on a $Rh_{201}$ Nanoparticle

We next conducted calculations on NO adsorption on different surface sites of a $Rh_{201}$ nanoparticle and compared our results with the DFT results of Nanba and Koyama [36]. We separately optimized the truncated octahedral $Rh_{201}$ nanoparticle and NO molecule. Furthermore, we modeled a single NO molecule adsorbed onto different on-top, bridge, and hollow sites of the $Rh_{201}$ (100) and (111) facets and at the ridge of both facets. The nomenclature used here is the same as the one used by Nanba and Koyama [36] and is explained in more detail in Figure S4 of the supporting information. Analogously to the study of Nanba and Koyama [36], we did not consider the bridge site $B_{32}$.

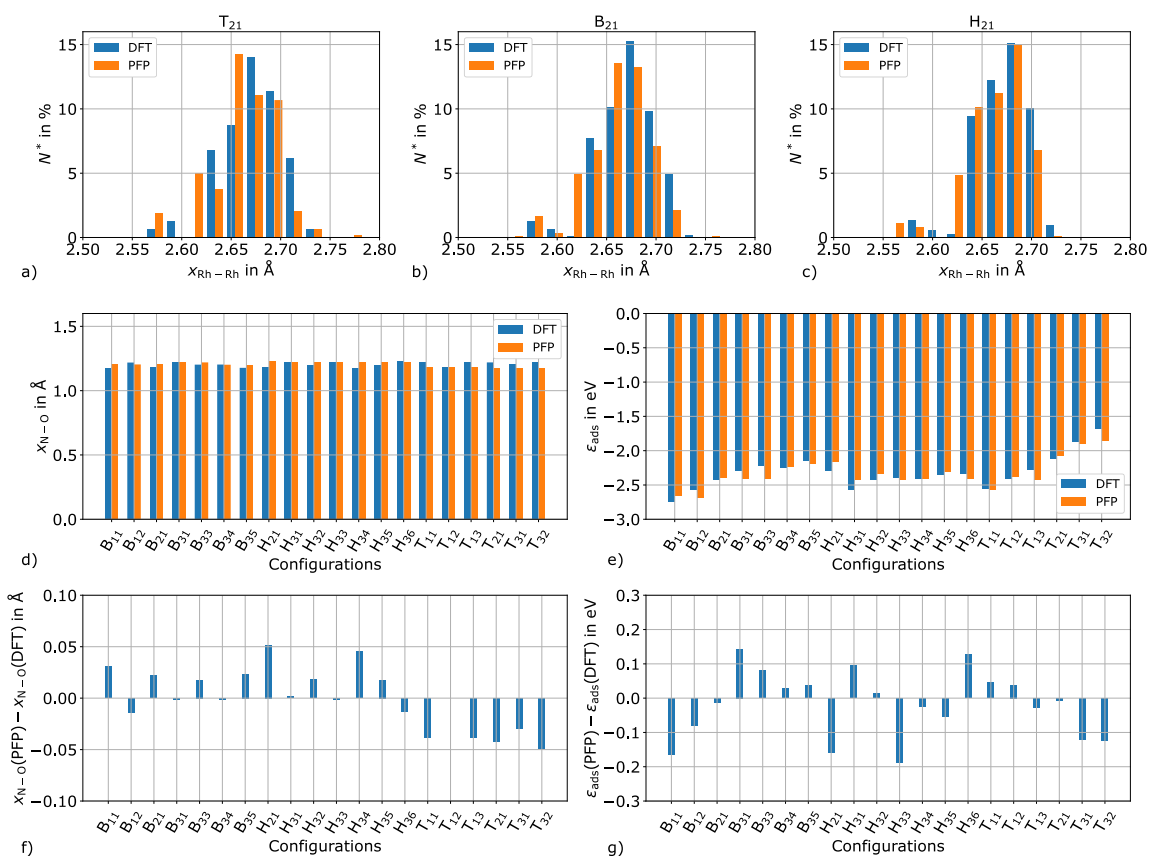

**Figure 3** Histograms for the nearest neighbor distances $x_{Rh-Rh}$ between the Rh atoms within the truncated octahedral $Rh_{201}$ nanoparticle calculated using the PFP and extracted from the DFT data used by Nanba and Koyama [36] with an NO molecule adsorbed at a) the on-top site ($T_{12}$), the bridge site ($B_{12}$), and the hollow site ($H_{12}$) of the (100) facet. The counts $N^*$ are given in percent. In the plot d), the values calculated for the N-O bond length $x_{N-O}$ using the PFP are compared with the DFT values calculated by Nanba and Koyama [36] for different configurations and plot f) contain the resulting absolute deviation of this comparison. In plot e), the values calculated using the PFP for the binding energy $\varepsilon_{ads}$ of the NO molecule on the $Rh_{201}$ nanoparticle are depicted together with the DFT values from Nanba and Koyama [36] and the resulting absolute deviations of this comparison are given in plot g).

We begin our discussion by analyzing the Rh nanoparticle itself. For this purpose, we extracted the nearest neighbor distances $x_{\text{Rh-Rh}}$ from the optimized configurations for the Rh atoms within the nanoparticles and built a histogram with a bin size of 0.015 Å. In Figures 3a, b, and c, the histograms resulting from the calculation using the PFP and from the DFT data used by Nanba and Koyama [36] in their study are compared exemplarily for the NO adsorbed on the on-top site, the bridge site and the hollow site of the (100) facet, i.e., $T_{12}$, $B_{12}$ and $H_{12}$. These histograms are representative of the histograms for all the other configurations. Due to the NO adsorbed molecule, the distribution of the Rh-Rh neighbor distances differed between the different configurations, as it can be taken from the examples in the depiction. To evaluate the similarity of the calculated histograms using the PFP with their DFT counterparts, we first compared the median value for $x_{\text{Rh-Rh}}$ averaged over all configurations and, second, we applied a simple method of Swain and Ballard [46] to quantify the intersection between both histograms. In the latter case, an intersection factor was calculated. If this factor is 0, the histograms are entirely different. If this factor is 1.0, the histograms are identical. The median values for $x_{\text{Rh-Rh}}$ for all configurations optimized using the PFP deviated just by 0.01178(24) Å in average from their DFT counterpart. Furthermore, the resulting intersection factors varied between 0.72 and 0.87. From both tests, we could conclude that the atomic position of the Rh atoms within the nanoparticle relaxed to a highly similar structure to the predicted with DFT.

In Figure 3d, the calculated bond length $x_{\text{N-O}}$ using the PFP is depicted in comparison to the DFT values of Nanba and Koyama [36]. The deviation from the calculated values for the N-O bond length for the bridge sites using PFP compared to the DFT values was lower than 0.033 Å. Most of the values showed an absolute deviation to the DFT data for the hollow sites lower than 0.018 Å, but for the $H_{21}$ and $H_{34}$ sites, the absolute deviations were 0.052 and 0.045 Å. The N-O bond length resulting from the calculation of the on-top sites were all close to the calculated value for the on-top site at the ridge ($T_{12}$), which was given with a high accuracy by less than 0.001 Å from the DFT value. This resulted in underestimated values compared to the respective DFT data, and the PFP could not effectively differentiate between the different on top-sites. However, the deviations were 0.05 Å.

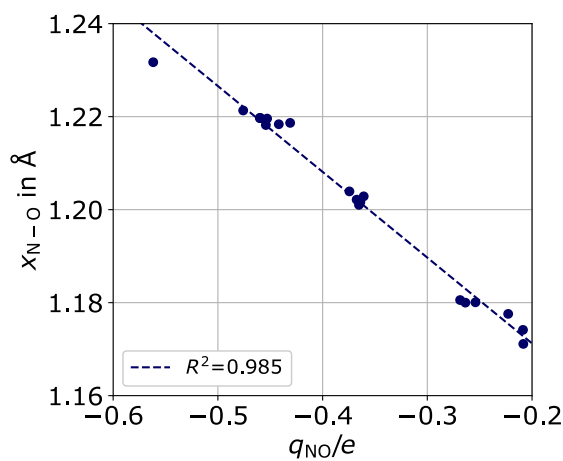

**Figure 4** Calculated N-O bond distance in dependency on the calculated charge of the NO molecules adsorbed at different bridge, on-top and hollow sites on a $Rh_{201}$ nanoparticle.

Besides, all values calculated with the PFP were longer than the bond length of the isolated NO of 1.168 Å. This behavior corresponds to the observed trend given by the DFT values. The adsorbed NO molecule forms a sigma bond with the nanoparticle and the sigma bond for the N-O bond becomes weaker, resulting in the N-O bond elongation. Simultaneously, the electrons are back-donated to the anti-bonding $\pi$ orbital of NO molecule leading to the bond elongation and negative charge of adsorbed NO molecule [47]. This can be typically seen by plotting the N-O bond distance against NO charge (Figure 4). A linear correlation between charge and N-O bond distance was observed.

In Figure 3e, the binding energy $\varepsilon_{ads}$ of the adsorbed NO molecule to the $Rh_{201}$ nanoparticle is compared with the DFT values of Nanba and Koyama [36]. The corresponding absolute deviations are given in Figure 3g. The values deviated only 0.78 eV on average, being the best approach for the hollow site $H_{33}$ on the (111) facet with a deviation of just 0.007 eV below its counterpart DFT value and the worst approach for the bridge site $B_{31}$ with an absolute deviation of 0.19 eV. This high deviation originated from the move of the NO molecule after geometry relaxation from the initial adsorption position at the bridge site $B_{31}$ in the (111) facet to the adjacent hollow site $H_{34}$. Unfortunately, this was also the case for the bridge site at the (111) facet. Except for those configurations, only six relaxed configurations showed a deviation between 0.1 eV and 0.16 eV. The absolute deviation of the other twelve configurations showed a deviation lower than 0.1 eV compared to the DFT values.

### 3.2 Application of the PFP

The validated PFP can be used to tackle problems that have been difficult by the DFT method. Here, we performed the adsorbate dynamics on nanoparticle surface and structural optimization of large systems of 2573 atoms.

Molecular Dynamics for Adsorption of NO on a $Rh_{201}$ Nanoparticle

Using the validated PFP, we conducted molecular dynamics simulations to analyze the adsorption dynamics at 298 K. It resulted in a computational time of 0.1 s per MD step. This relatively slow iteration time originated from using a regular GPU device without any parallelization.

In Table 1, we compared the NO molecule site after optimization and during the production run of the simulation. As it can be taken from the data, only the on-top site $T_{11}$ at the ridge was stable at 298 K. The NO molecule adsorbed to the other on-top sites at ridge $T_{12}$ and $T_{13}$ diffused to their adjacent bridge sites $B_{11}$ and $B_{12}$, respectively. This was also the case for the NO molecule adsorbed on the on-top sites $T_{21}$ and $T_{32}$ at the (100) facet and the (111) facet diffusing to the adjacent bridge sites $B_{21}$ and $B_{33}$ in the respective facet. After the adsorption at the bridge site $B_{21}$, the NO molecule adopted an orientation parallel to the (100) facet pointing perpendicularly to the bridge built by the Rh atoms. Moreover, the on-top site $T_{31}$ diffused to the adjacent hollow site $H_{33}$, both in the (111) facet. As it can be seen from Figure 3, the bridge sites and the hollow site $H_{33}$ showed lower adsorption energy than the on-top sites where the NO molecule is initially adsorbed. The adsorption energy for the on-top site $T_{11}$ was one of the lowest among all adsorption states. Thus, the NO molecule remained attached at this site. These observations were statistically confirmed with the calculated bond length of the NO molecule, which is characteristic to the respective adsorption site. Only the value $x_{N-O}^{298\,K}$ at 298 K for the on-top site $T_{11}$ at ridge was closer to its corresponding optimization value. The calculated values for the NO molecule initially adsorbed on the remaining on-top site were in accordance with the values for the more stable bridge and the hollow sites. The NO molecule initially adsorbed at all bridge sites remained in its initial adsorption site and the NO molecule adsorbed on the bridge site $B_{21}$ at the (100) facet also showed the same orientational behavior as described before.

Finally, only the NO molecule adsorbed at the hollow sites $H_{33}$ and $H_{36}$ at the (111) facet remained adsorbed in their initial position, whereas, for the remaining hollow sites, the NO molecule diffused to the more stable nearest ridge sites. Remarkable is that the NO molecule initially positioned on the site $H_{31}$ at the (111) facet diffused first to the $H_{34}$ site on the same facet to finally hopped to the ridge $B_{12}$ site. All these observations agreed with the results for the adsorption energy and were statistically supported by the values provided for the N-O bond lengths.

**Table 1** Dynamics of the adsorbed NO starting from the optimization site. $x_{N-O}^{opt}$ is the N-O bond length calculated after optimization and $x_{N-O}^{298\,K}$ is the N-O bond length calculated at 298 K

| Site (opt.) | Site (298 K) | $x_{N-O}^{opt.}$ in Å | $x_{N-O}^{298\,K}$ in Å |
|---|---|---|---|
| ridge $T_{11}$ | | 1.18057 | 1.199(12) |
| ridge $T_{12}$ | ridge $B_{11}$ | 1.18007 | 1.2142(56) |
| ridge $T_{13}$ | ridge $B_{12}$ | 1.18001 | 1.20744(79) |
| (100) facet $T_{21}$ | (100) facet $B_{21}$[a] | 1.17759 | 1.242(22) |
| (111) facet $T_{31}$ | (111) facet $H_{33}$ | 1.17415 | 1.22023(76) |
| (111) facet $T_{32}$ | (111) facet $B_{34}$ | 1.17112 | 1.21925(47) |
| ridge $B_{11}$ | | 1.20214 | 1.2195(41) |
| ridge $B_{12}$ | | 1.20393 | 1.20593(54) |
| (100) facet $B_{21}$[a] | | 1.20287 | 1.3122(48) |
| (111) facet $B_{34}$ | | 1.20160 | 1.21915(35) |
| (111) facet $B_{35}$ | | 1.20104 | 1.21954(55) |
| (100) facet $H_{21}$ | ridge $B_{11}$ | 1.23171 | 1.20587(93) |
| (111) facet $H_{31}$ | ridge $B_{11}$ | 1.22133 | 1.2118(62) |
| (111) facet $H_{32}$ | ridge $B_{12}$[b] | 1.21958 | 1.2088(14) |
| (111) facet $H_{33}$ | | 1.21819 | 1.21961(80) |
| (111) facet $H_{34}$ | ridge $B_{12}$ | 1.21970 | 1.20641(45) |
| (111) facet $H_{35}$ | facet $H_{33}$ | 1.21865 | 1.21944(50) |
| (111) facet $H_{36}$ | | 1.21837 | 1.21978(74) |

a. Orientated parallelly to the (100) facet for $t > t_{equil}$. b. Transition over $H_{34}$

Pt nanoparticles supported on SnO$_2$

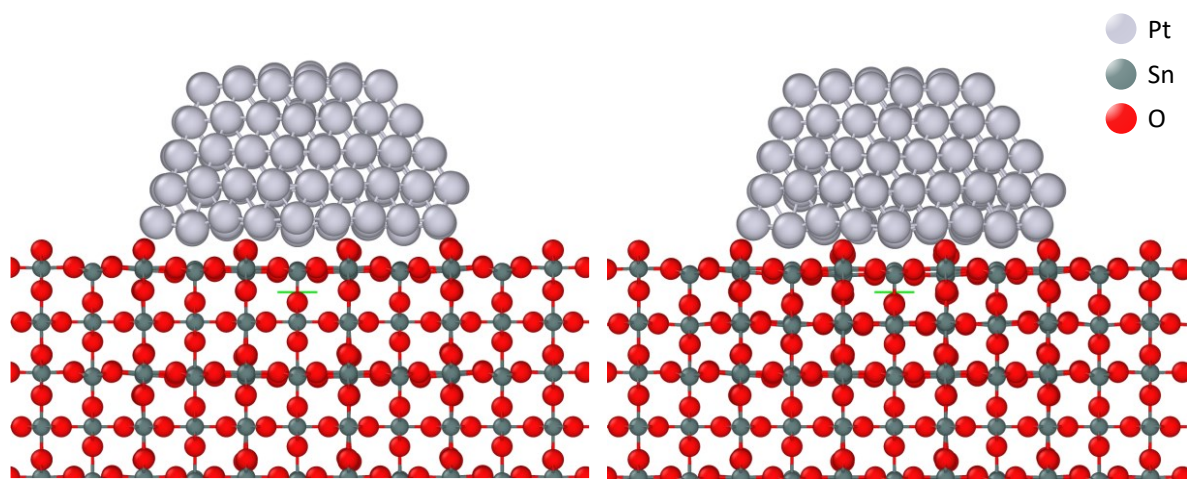

**Figure 5** Snapshots of the Pt-SnO$_2$ system showing two adjacent (100) and (111) Pt facets a) before and b) after optimization. The view on the SnO$_2$ (001) plane is depicted for both cases. The images were visualized using OVITO [48].

The structure of Pt nanoparticles supported on SnO$_2$ was calculated by DFT in a previous study [16]. However, they considered only two layers of SnO$_2$ as the slab model due to the extremely expensive computational cost. We considered five layers of SnO$_2$ as the slab model to make the slab model closer to reality. The Pt nanoparticles were constructed from 233 atoms (2.23 nm), and the total number of atoms in the system was 2573. While only the bottom three layers were fixed, the structure was optimized by the FIRE algorithm [43] with a force tolerance of 0.005 eV/Å (Fig. 5). Because this is the largest system studied here, it is worth exemplarily to report the simulation time for optimization. A single calculation step took about 0.125 s and less than 16 GB of GPU memory, showing that large-scale structural optimization could be performed quickly using a single regular GPU device.

Pt nanoparticles supported on SnO$_2$ were distorted relative to their structure in vacuum (see Figure 5). This is in qualitative agreement with the experimental results using STEM [49]. In addition, the Bader charges calculated by PFP were positive at the triple-phase boundary (see Figure 6), especially for the Pt atoms at the oxygen SnO$_2$ sites, showing the same trend as in previous DFT studies [16]. These results are essential for understanding the effect of the carrier on the reaction on the nanoparticles and indicate that PFP can be used to explore the supporting materials.

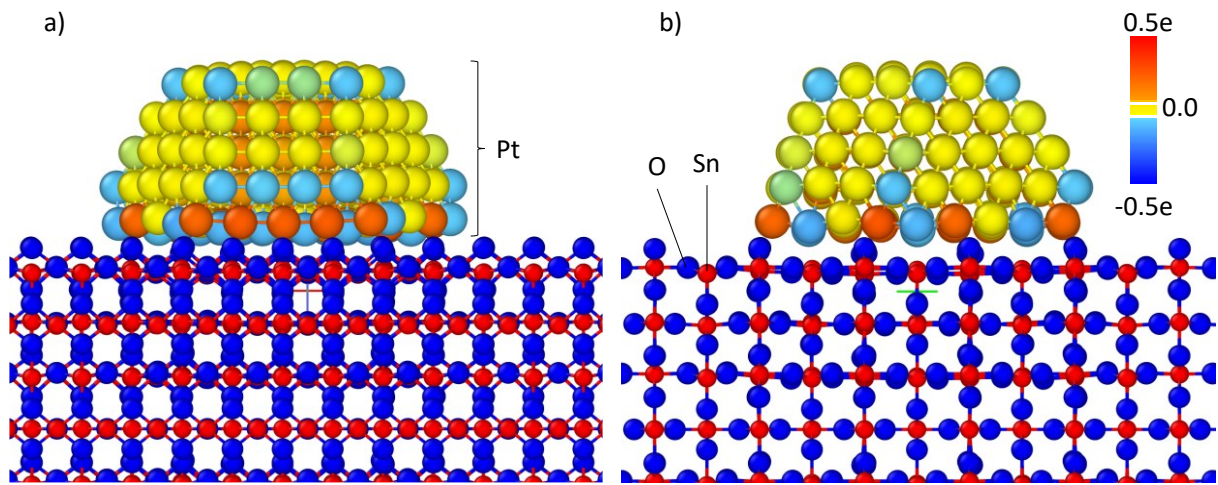

**Figure 6** Calculated charge distribution after optimization on the supported Pt nanoparticle, whereas the view a) on the SnO$_2$ (1$\bar{1}$0) plane shows a (100) Pt facet and the view on b) the SnO$_2$ (001) plane shows two adjacent (100) and (111) Pt-facets. The images were visualized using OVITO [48].

4. **Summary and Conclusions**

In this study, we validated a novel universal Neural Network potential, PFP, to simulate real-system nanoparticles and conducted further simulations that are difficult to tackle with the state-of-the-art DFT methods for the description of nanoparticles as catalysts.

The current research in the computational design of catalysts requires new techniques that deliver accurate results comparable to DFT methods, enabling the computation of systems that are presently difficult to tackle with this method, such as large nanoparticle systems or dynamics under a finite temperature. The PFP is a Graph NNP with an implemented E(3) invariance. The PFP is an ideal candidate to assist the computational design of materials, as it was trained to describe a broad palette of materials, including metals and molecules.

The PFP was validated in this study with DFT data for the cohesive energy for Ru monometallic nanoparticles, for the excess energy of truncated octahedral (PdRuCu)$_{201}$ ternary alloy nanoparticles, and the adsorption of NO in Rh$_{201}$ nanoparticles. The accuracy in reproducing DFT values for the cohesive energy of Ru nanoparticles using the PFP was relatively high, with an absolute deviation of 0.0615(79) eV in average. Furthermore, the progression of the cohesive energy depending on the nanoparticle size resulting from the calculation with the PFP could be taken as equivalent to the DFT results, which was not the case for the calculated progression using an already established EAM potential. The PFP could reproduce the excess energy for the (PdRuCu)$_{201}$ ternary alloy nanoparticles deviating lower than 0.011 eV from the reference DFT values. Moreover, the PFP could describe all the Rh$_{201}$-NO interfacial system constituents, i.e., the Rh nanoparticle, the adsorbed NO molecule, and the interfacial interaction with accuracy like DFT calculations.

Finally, using the PFP, MD simulations were performed on the Rh$_{201}$-NO interfacial system at 298 K and to study a large-scale Pt nanoparticle supported on SnO$_2$. We identified all bridge sites, the on-top site T$_{11}$ at the ridge, and the H$_{33}$ and the H$_{36}$ hollow sites at the (111) facet to be the most stable adsorption sites for the Rh$_{201}$-NO interfacial system. This behavior differed from the optimization sites predicted at zero Kelvin. On the other hand, the optimization of a supported Pt nanoparticle system on SnO$_2$ with 2573 atoms was possible using the PFP with outstanding computational efficiency compared to DFT calculations. The optimization delivered a distorted Pt nanoparticle structure in qualitative agreement with reported experimental observations and distribution of Bader charges with the same trend as previous DFT studies. Further experimental study is desired to compare the results from large-scale PFP simulations.

# Supporting Information

## Calculations of Real-System Nanoparticles Using Universal Neural Network Potential PFP

Gerardo Valadez Huerta[a,*], Yusuke Nanba[a], Iori Kurata[c,*], Kosuke Nakago[c], So Takamoto[c], Chikashi Shinagawa[c], Michihisa Koyama[a,b,*],

[a] Research Initiative for Supra Materials, Shinshu University

[b] Open Innovation Institute, Kyoto University

[c] Preferred Networks, Inc.

*valadez@shinshu-u.ac.jp;  kurata@preferred.jp;  koyama_michihisa@shinshu-u.ac.jp, koyama.michihisa.5k@kyoto-u.ac.jp

Ru Nanoparticles

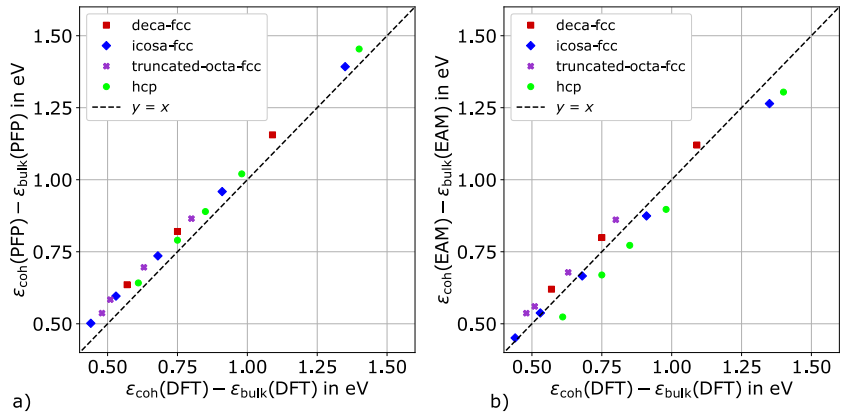

**Figure S1** Calculated difference $\varepsilon_{coh} - \varepsilon_{bulk}$ between the cohesive energy $\varepsilon_{coh}$ and the total energy per atom of the bulk phase $\varepsilon_{bulk}$ for the a) 5-fold twinned decahedral fcc (in red), b) icosahedral fcc (in blue), c) truncated-octahedral fcc (in purple) and the hcp (in green) Ru structures in dependency on the corresponding DFT values taken from the study of Nanba et al.[S1]. a) Values calculated using the PFP. b) Values calculated using the EAM potential of Fortini et al.[S2].

To compare absolute energies is not helpful. For that reason, we summarize in Figure S1a and b the calculated values for the difference $\varepsilon_{coh} - \varepsilon_{bulk}$ between the cohesive energy $\varepsilon_{coh}$ and the total energy per atom of the respective bulk phase $\varepsilon_{bulk}$ using the PFP and the EAM potential, respectively. We plotted the values over their DFT counterparts from the study of Nanba *et al.*[S1] differentiating between the mentioned nanoparticle structures. The absolute deviation (averaged over all nanoparticle structures) to the DFT values was of 0.0615(79) eV and 0.052(12) eV for the PFP and the EAM potential, respectively. One can argue that the values calculated by both classical approaches show a similar accuracy. However, the values calculated with the PFP were closer to the identity line than the values calculated with the EAM potential (see Figure S1), which is statistically supported by the uncertainty of the mean values for the absolute deviations. The calculated values with the PFP and the EAM increased parallelly to the DFT values for most cases but for the values calculated with the EAM potential for the icosahedral fcc structure, where the progression went from higher values than their DFT counterparts to smaller values as the cohesive energy increases. It resulted in a slope calculated from a linear regression of 1.0059(22) for the values calculated using PFP and 0.917(32) for the values calculated with EAM with an $R^2$ value of 0.993 and 0.981 respectively.

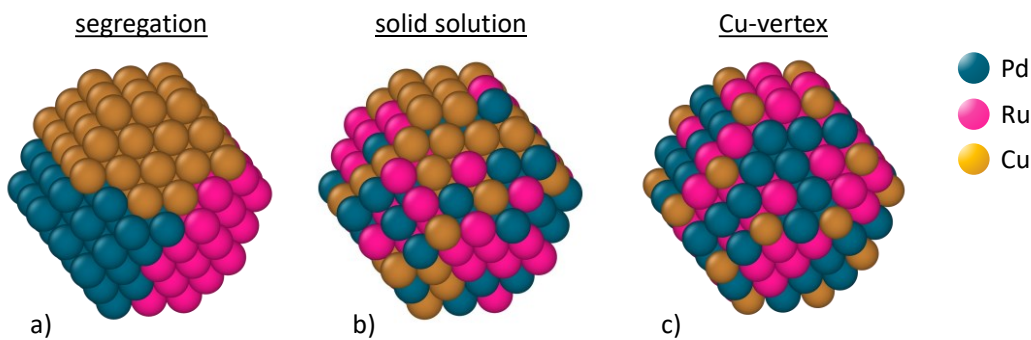

**Figure S2** Example a) segregation (pure clusters for each component) and b) solid solution (complete randomness) structures for the truncated octahedral $Pd_{67}Ru_{67}Cu_{67}$ as well as example c) Cu-vertex (with all vertices occupied by a Cu atom) for the truncated octahedral $Pd_{90}Ru_{90}Cu_{21}$ as illustration for the configurations investigated in this study. The images were visualized using OVITO [S3].

**Table S1** Mean values for the pairwise multicomponent short-range order (SRO) $\alpha_{ij}$ (i, j = Pd, Ru, Cu) for different ternary alloy truncated octahedral $Pd_zRu_zCu_{201-2z}$ nanoparticle configurations (solid solution (SS), segregation (SG), Cu-vertex (VT)).

| $Pd_zRu_zCu_{201-2z}$ | | j = Pd | j = Ru | j = Cu |
|---|---|---|---|---|
| z = 67 | SS | -0.049(23) | 0.019(11) | 0.030(16) |
|  | SG | -1.4360 | 0.7299 | 0.7062 |
| z = 80 | SS | 0.00262(69) | -0.0001(11) | -0.0049(18) |
|  | SG | -1.0860 | 0.7534 | 0.6489 |
| z = 90 | SS | -0.020(12) | 0.020(10) | 0.000(12) |
|  | VT | -0.0841(24) | -0.00120(46) | 0.3165(86) |

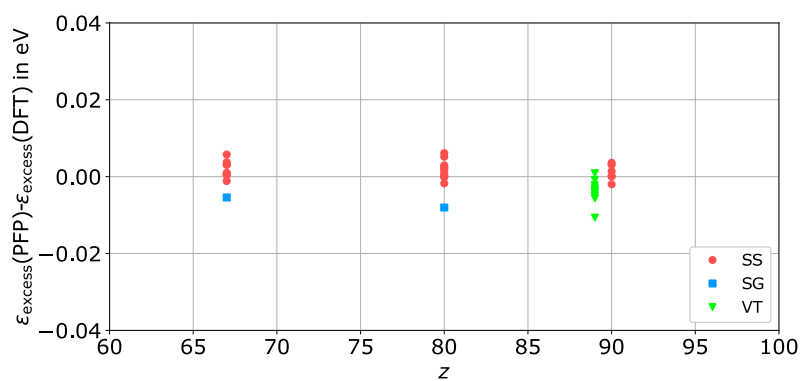

**Figure S3** Absolute deviation for the excess energy $\varepsilon_{excess}$ between the calculated values using PFP and DFT for different $Pd_zRu_zCu_{201-2z}$ nanoparticles in dependency on the variable $z$. Red circles: solid solution (SS). Green triangles: Cu-vertex (VT). Blue squares: segregation (SG).

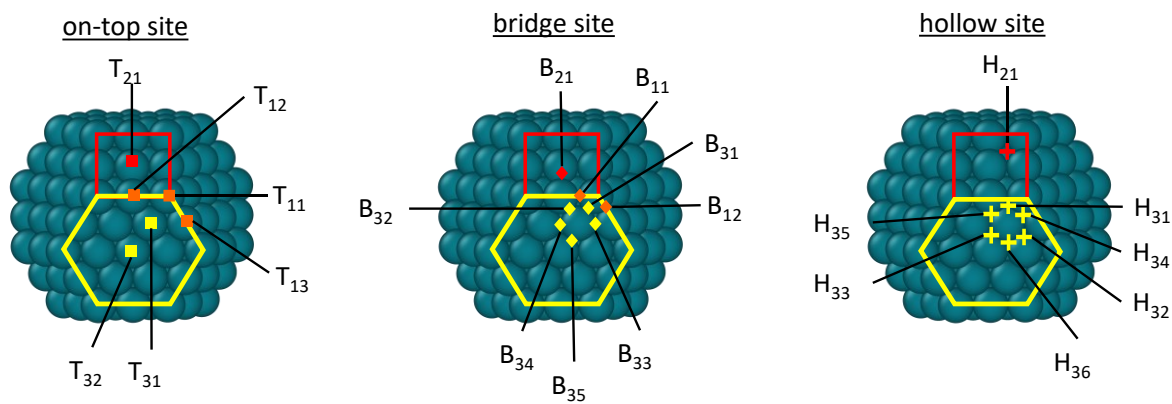

**Figure S4** NO adsorption sites on the truncated octahedral Rh$_{201}$ nanoparticle. The surface within the red rectangle corresponds to the (100) facet and the surface within the yellow hexagon the (111) facet. The adsorption sites in ridge (T$_{1X}$, B$_{1X}$), on the (100) facet (T$_{2X}$, B$_{2X}$, H$_{2X}$) and on the (111) facet (T$_{3X}$, B$_{3X}$, H$_{3X}$) are represented by orange, red and yellow symbols respectively. The images were visualized using OVITO [S3].